\documentclass[aps,pra,preprint,showpacs,eqsecnum,floatfix]{revtex4-2}

\usepackage{braket}
\usepackage{amsmath}
\usepackage{amssymb}

\begin{document}

%=================================================================
% Full title of the paper (Capitalized)
\title{Hidden rotation symmetry of the Jordan-Wigner transformation and its application to measurement in quantum computation}

% Authors, for the paper (add full first names)
\author{Grant Davis} 
\affiliation{%
Department of Physics, Georgetown University, 37th and O Sts. NW, Washington, DC 20057, USA}

\author{James K Freericks }
\affiliation{%
Department of Physics, Georgetown University, 37th and O Sts. NW, Washington, DC 20057, USA}

% Abstract (Do not insert blank lines, i.e. \\) 
\begin{abstract}
Using a global rotation by $\theta$ about the $z$-axis in the spin sector of the Jordan-Wigner transformation rotates Pauli matrices $X$ and $Y$ in the $x-y$-plane, while it adds a global complex phase to fermionic quantum states that have a fixed number of particles. With the right choice of angles, this relates expectation values of Pauli strings containing products of $X$ and $Y$ to different products, which can be employed to reduce the number of measurements needed when simulating fermionic systems on a quantum computer. Here, we derive this symmetry and show how it can be applied to systems in Physics and Chemistry that involve Hamiltonians with only single-particle (hopping) and two-particle (interaction) terms. We also discuss the consequences of this for finding efficient measurement circuits in variational ground state preparation.
\end{abstract}

\date{\today}
\maketitle

%%%%%%%%%%%%%%%%%%%%%%%%%%%%%%%%%%%%%%%%%%

%%%%%%%%%%%%%%%%%%%%%%%%%%%%%%%%%%%%%%%%%%
%\setcounter{section}{-1} %% Remove this when starting to work on the template.
\section{Introduction}

One of the most important applications of quantum computing in physical science is the problem of preparing the ground state of a physical Hamiltonian. While many different algorithms have been proposed for this problem, one of the most common approaches is variational algorithms, exemplified by the variational quantum eigensolver (VQE) ~\cite{peruzzo_variational_2014}. Here, a quantum state ansatz is used with a set of parameters employed to construct the state, and these parameters are then optimized by minimizing the energy. A key element of this algorithm is the need to measure both the Hamiltonian and the derivatives for how the expectation value of the Hamiltonian changes as a function of the parameters used in the ansatz. Even with a shallow state-preparation circuit, the number of measurements required to carry out this work can be enormous. This leads to the important problem of making the measurement step (of the Hamiltonian and its derivatives) as efficient as possible.

The problem has been studied for a number of years now and a recent review article has appeared~\cite{patel_quantum_2025}. One of the approaches proposed is to separate the Hamiltonian into fragments, and compute the averages of each piece, eventually combining all of the fragments to determine the expectation value~\cite{jena2019paulipartitioningrespectgate}. By creating circuits that measure commuting operators at the same time, one can make the measurements more efficient. It is sometimes possible to even take nominally noncommuting operators and map them to a new basis where the relevant operators commute (an example of that is to look at a tight-binding Hamiltonian, where elements do not commute in the position basis, but after diagonalizing to the band structure multiplied by number operators, each element commutes)~\cite{patel_quantum_2025}. 

A second approach uses classical shadows~\cite{huang_predicting_2020}. Here, a set of typically randomized measurements are made by randomizing the axis for the measurement on each qubit, and the results of those measurements are employed to construct an approximation of the density matrix, which then can be employed to measure all observables. 

A third approach is to analyze the measurement in terms of an underlying Lie algebraic structure in the Hamiltonian~\cite{PRXQuantum.2.040320, Izmaylov_2021}. If the Lie algebra can be kept relatively small in size, then one can use group theoretic methods to reduce the number of measurements needed for many expectation values.

Here, we focus on the fragment method and describe a number of aspects related to it. In particular, the fragments we work with will involve the Hermitian combination of quadratic and quartic terms (same number of creation and annihilation operators in each product) from the Hamiltonian. These terms are modified when one performs the Jordan-Wigner transformation~\cite{jordan-wigner, lieb-shultz-mattis} to convert from a fermionic system to a spin system, which can be more directly simulated on conventional quantum computers.

The strategy used for preparing the fermionic system for calculations on a quantum computer involves a few steps. First, because spin one-half states have only two states, as do fermions, we map the states according to the electron number, corresponding to $\ket{0}$ and $\ket{1}$. Second, nothing needs to be done with respect to the up-spin terms versus the down spin terms, since we can have the fermions for up and down spins commute with each other, as long as we keep the number of up spin fermions and down-spin fermions each conserved in our quantum state~\cite{lieb-two-theorems}. Third, to guarantee anticommutation between the fermions at any two lattice sites, we need to introduce a Jordan-Wigner string~\cite{jordan-wigner}, which changes the commutation relations to anticommutation relations for the spin operators of the qubits, which represent electronic operators of the same spin. This completes the mapping, whose details are discussed further below.

The paper is organized as follows. In Sec.~2, we describe the details of the Jordan-Wigner transformation and prove the hidden symmetry result. In Sec.~3, we apply it to quadratic operators, while in Sec.~4, it is applied to quartic operators. We consider how this will be used in Physics in Sec.~5 and in Chemistry in Sec.~6. Conclusions follow in Sec.~7.

\section{Materials and Methods}

The methods used for this paper rely on exploiting the hidden symmetry of the Jordan-Wigner transformation, which is discussed next.

\subsection{Hidden Symmetry of the Jordan-Wigner Transformation}

The fermionic system is assumed to be discretized onto a lattice. In Physics problems, it is often a real spatial lattice, while in Chemistry problems, it is an abstract ``lattice'' constructed from the different spatial orbitals used to describe the ``active space'' of the molecule. No matter what lattice the system is on, we choose a specific ordering of the lattice sites, so that we can label the sites from $0$ to $N-1$ in order. The fermionic creation $\hat{c}_{i\sigma}^\dagger$ and annihilation $\hat{c}_{i\sigma}^{\phantom{\dagger}}$ operators satisfy the following anticommutation relations
\begin{equation}
    \{\hat{c}_{i\sigma}^{\phantom{\dagger}},\hat{c}_{j\sigma'}^\dagger\}=0,~~\{\hat{c}_{i\sigma}^{\phantom{\dagger}},\hat{c}_{j\sigma'}^{\phantom{\dagger}}\}=0,~~\text{and}~~\{\hat{c}_{i\sigma}^\dagger,\hat{c}_{j\sigma'}^{\phantom{\sigma}}\}=\delta_{ij}\delta_{\sigma\sigma'}.
\end{equation}
Here, the anticommutator is $\{\hat{A},\hat{B}\}=\hat{A}\hat{B}+\hat{B}\hat{A}$ and $\sigma$ denotes the spin state of the fermion (either $\uparrow$ or $\downarrow$ along the $z$-axis).

The spin operators are denoted as $\mathbb{I}$, $\hat{X}$, $\hat{Y}$, and $\hat{Z}$, for each lattice site and spin degree of freedom of the fermion. Here, we have the matrix representation of each operator is given by
\begin{align}
    \mathbb{I}\leftrightarrow\begin{pmatrix}1&0\\0&1\end{pmatrix}, ~~\hat{X}\leftrightarrow\begin{pmatrix}0&1\\1&0\end{pmatrix},~~\hat{Y}\leftrightarrow\begin{pmatrix}0&-i\\i&\phantom{-}0\end{pmatrix},~~\text{and}~~\hat{Z}\leftrightarrow\begin{pmatrix}1&\phantom{-}0\\0&-1\end{pmatrix},
\end{align}
which are just the familiar Pauli spin matrices. Note that we are going between abstract operators and their matrix representation without explicitly converting from one to the other. This should not cause any problems in the way the discussion is being given here. Note further that the Pauli spin matrices satisfy $\hat{X}\hat{Y}=i\hat{Z}$ and $\hat{X}\hat{Z}+\hat{Z}\hat{X}=0$ and similarly for other choices, as is ordinarily the rule for angular momentum in the spin one-half representation.

The qubits are chosen here to run sequentially from $0$ to $N-1$ for the fermionic operators labeled $0\uparrow$ to $N-1\uparrow$, while the qubits labeled from $N$ to $2N-1$ correspond to the down spin fermionic operators ranging over the same ordered spatial lattice sites as we used for the up spin (now with the spatial index equal to $i-N$).
The key point about these spin operators is that they commute, rather than anticommute when on different sites, while they anticommute on the same site. The Jordan-Wigner transformation makes them anticommute on different lattice sites, as required to represent fermionic states.

We now describe in detail how the Jordan-Wigner transformation works. Since the $\ket{0}$ state is the $\ket{\uparrow}$ state, and the $\ket{1}$ state is the $\ket{\downarrow}$ state, we find the fermionic creation operator is mapped to the spin lowering operator. We are going to describe the transformation for the spin-up operators only. A similar transformation for the spin-down operators will also follow. The claim is that we have 
\begin{equation}
    \hat{c}_{i\uparrow}^\dagger=\frac{1}{2}\prod_{j=0}^{i-1}e^{i\frac{\pi}{2}(\mathbb{I}_j-\hat{Z}_j)}\hat{\sigma}_i^-~~\text{and}~~\hat{c}_{i\uparrow}^{\phantom{\dagger}}=\frac{1}{2}\hat{\sigma}_i^+\prod_{j=0}^{i-1}e^{-i\frac{\pi}{2}(\mathbb{I}_j-\hat{Z}_j)},
\end{equation}
where $\hat{\sigma}^\pm=\hat{X}\pm i\hat{Y}$. 

One can directly verify that these operators anticommute between different sites, while at the same site, exponential factors cancel, and the anticommutation relations hold. As an example, we have for $i>k$,
\begin{align}
    \{\hat{c}_{i\uparrow}^\dagger,\hat{c}_{k\uparrow}^{\phantom{\dagger}}\}&=\frac{1}{4}\left (\prod_{j=k}^{i-1}e^{i\frac{\pi}{2}(\mathbb{I}_j-\hat{Z}_j)}\hat{\sigma}_i^-\hat{\sigma}_k^++\hat{\sigma}_k^+\prod_{j=k}^{i-1}e^{i\frac{\pi}{2}(\mathbb{I}_j-\hat{Z}_j)}\hat{\sigma}_i^-\right )\nonumber\\
    &=\frac{1}{4}\prod_{j=k+1}^{i-1}e^{i\frac{\pi}{2}(\mathbb{I}_j-\hat{Z}_j)}\left (  \hat{\sigma}_i^-\hat{Z}_k\hat{\sigma}_k^++\hat{\sigma}_k^+\hat{Z}_k\hat{\sigma}_i^- \right )   \nonumber\\
    &=\frac{1}{4}\prod_{j=k}^{i-1}e^{i\frac{\pi}{2}(\mathbb{I}_j-\hat{Z}_j)}\left (\hat{\sigma}_i^-\hat{\sigma}_k^+-\hat{\sigma}_k^+\hat{\sigma}_i^-\right )=0.
\end{align}
Here, we used the facts that $e^{\pm i\frac{\pi}{2}\hat{Z}}=\pm i \hat{Z}$ and that $\hat{Z}$ anticommutes with $\hat{X}$ and $\hat{Y}$.
Note how the Jordan-Wigner string, given by the product of exponential operators from $k$ to $i-1$, changes the sign of the anticommutator to make it a commutator.
The transformation for the down spin fermions follows in the same way, except that the Jordan-Wigner string runs from $j=N$ to $j=N+i-1$, because we are working with a representation where the up and down spin fermionic operators commute, as opposed to anticommute.

Now, we discuss the hidden symmetry of the system in the spin representation. Consider the global rotation by the angle $\theta$ defined by the following unitary operator:
\begin{equation}
    \hat{R}=\prod_{j=0}^{2N-1}e^{-i\frac{\theta}{2}\hat{Z}_j}.
\end{equation}
Then we have
\begin{equation}
    \hat{R} \hat{X}_j\hat{R}^\dagger=\cos\theta\hat{X}_j+\sin\theta \hat{Y}_j,~~\hat{R}\hat{Y}_j\hat{R}^\dagger=-\sin\theta\hat{X}_j+\cos\theta\hat{Y}_j,~~\text{and}~~\hat{R}\hat{Z}_j\hat{R}^\dagger=\hat{Z}_j.
\end{equation}
For example, since $\hat{R}=\prod_{j=0}^{2N-1}\big(\cos\tfrac{\theta}{2}\mathbb{I}-i\sin\tfrac{\theta}{2}\hat{Z}_j\big)$, we have 
\begin{align}
    \hat{R}\hat{X}_j\hat{R}^\dagger&=(\cos\tfrac{\theta}{2}\mathbb{I}_j-i\sin\tfrac{\theta}{2}\hat{Z}_j)\hat{X}_j(\cos\tfrac{\theta}{2}\mathbb{I}+i\sin\tfrac{\theta}{2}\hat{Z}_j)\nonumber\\
    &=\cos^2\tfrac{\theta}{2}\hat{X}_j+i\cos\tfrac{\theta}{2}\sin\tfrac{\theta}{2}(-\hat{Z}_j\hat{X}_j+\hat{X}_j\hat{Z}_j)+\sin^2\tfrac{\theta}{2}\hat{Z}_j\hat{X}_j\hat{Z}_j\nonumber\\
    &=\cos\theta\hat{X}_j+\sin\theta \hat{Y}_j.
\end{align}
The fermionic creation operator then transforms as $\hat{c}_{j\sigma}^\dagger\to e^{i\theta}\hat{c}_{j\sigma}^\dagger$, because $\hat{\sigma}^-_j=\hat{X}_j-i\hat{Y}_j$ and
\begin{equation}
    \hat{X}_j-i\hat{Y}_j\to\cos\theta\hat{X}_j+\sin\theta\hat{Y}_j+i\sin\theta\hat{X}_j-i\cos\theta\hat{Y}_j=e^{i\theta}(\hat{X}_j-i\hat{Y}_j).
\end{equation}
Then, if we act on a fermionic state $\ket{\psi}$ that has exactly $M$ electrons, we find that $\hat{R}\ket{\psi}=e^{iM\theta}\ket{\psi}$ and $\bra{\psi}\hat{R}^\dagger=\bra{\psi}e^{-iM\theta}$, so that
\begin{equation}
    \braket{\psi|\hat{O}|\psi}=\braket{\psi|\hat{R}^\dagger\hat{R}\hat{O}\hat{R}^\dagger\hat{R}|\psi}=\braket{\psi|\hat{R}\hat{O}\hat{R}^\dagger|\psi},
\end{equation}
because the global phases from the bra and ket cancel. First consider the case where $\theta=\tfrac{\pi}{2}$. Then the rotation maps $\hat{X}\to \hat{Y}$, $\hat{Y}\to -\hat{X}$ and $\hat{Z}\to \hat{Z}$. We find that the expectation value of a Pauli string that involves a product of $\hat{X}$, $\hat{Y}$, and $\hat{Z}$ operators (including the Jordan-Wigner strings), is equal to the expectation value of the same operator with $\hat{X}\to \hat{Y}$, $\hat{Y}\to -\hat{X}$, and $\hat{Z}\to \hat{Z}$. This allows us then to relate the expectation value of a string of $\hat{X}$ and $\hat{Y}$ operators (with Jordan-Wigner strings) to be mapped to plus or minus the complementary string, with $\hat{X}$ and $Y$ interchanged, and a plus sign for an even number of $\hat{Y}$ operators in the original operator and a minus sign for an odd number. This is the hidden symmetry. It allows us to reduce the number of measurements required for measuring expectation values of products of fermionic operators by at least a factor of two, when a nonentangling measuring scheme is used for those operators. We will consider other angles as well below, which allow us to reduce the number of measurement circuits by a factor of four.

Note that the U(1) symmetry relating the global phase of a fermionic quantum state to the number of fermions is a well-known symmetry. It is the use of this symmetry, via the Jordan-Wigner transformation, and how it affects matrix elements of Pauli strings that is the ``hidden symmetry'' that we work with here.
We will describe different measurement schemes that use this symmetry in the next two subsections.

\section{Results}

\subsection{Application to Quadratic Operators}

A typical Hermitian operator that conserves the total particle number and is quadratic in the operators takes the form
\begin{equation}
    \hat{T}_{ik\sigma}=\hat{c}_{i\sigma}^\dagger\hat{c}_{k\sigma}^{\phantom{\dagger}}+\hat{c}_{k\sigma}^\dagger\hat{c}_{i\sigma}^{\phantom{\dagger}}.
\end{equation}
For concreteness, we will work with the spin-up case. After the Jordan-Wigner transformation, this becomes
\begin{align}
    \hat{T}_{ik\uparrow}&\rightarrow \frac{1}{4}\prod_{j=k}^{i-1}e^{-i\frac{\pi}{2}(\mathbb{I}_j-\hat{Z}_j)}\hat{\sigma}_i^-\hat{\sigma}_k^++\frac{1}{4}\hat{\sigma}_k^-\hat{\sigma}_i^+\prod_{j=k}^{i-1}e^{i\frac{\pi}{2}(\mathbb{I}_j-\hat{Z}_j)}\nonumber\\
    &=\frac{1}{4}\left (\prod_{j=k+1}^{i-1}e^{-i\frac{\pi}{2}(\mathbb{I}_j-\hat{Z}_j)} \hat{\sigma}_i^-\hat{Z}_k\hat{\sigma}_k^++\hat{\sigma}_k^-\hat{Z}_k\hat{\sigma}_i^+\prod_{j=k+1}^{i-1}e^{i\frac{\pi}{2}(\mathbb{I}_j-\hat{Z}_j)}\right )\nonumber\\
    &=\frac{1}{4}\left (\prod_{j=k+1}^{i-1}e^{-i\frac{\pi}{2}(\mathbb{I}_j-\hat{Z}_j)} \hat{\sigma}_i^-\hat{\sigma}_k^++\hat{\sigma}_k^-\hat{\sigma}_i^+\prod_{j=k+1}^{i-1}e^{i\frac{\pi}{2}(\mathbb{I}_j-\hat{Z}_j)}\right ).
\end{align}
where we assumed $i>k$. Note that we also used the facts that $e^{\pm i\frac{\pi}{2}\hat{Z}}=\pm i\hat{Z}$, $\hat{Z}\hat{\sigma}^+=\hat{\sigma}^+$ and $\hat{\sigma}^-\hat{Z}=\hat{\sigma}^-$. Note further that when the two terms are nearest neighbors, so $k=i-1$, then there are no Jordan-Wigner strings. Furthermore, the Jordan-Wigner strings can both be written as $\prod_{j=k+1}^{i-1}\hat{Z}_j$ after taking into account the matrix term and the constant term.

Expanding in terms of $\hat{X}$ and $\hat{Y}$ matrices, we see that $\hat{\sigma}^-_i\hat{\sigma}^+_k+\hat{\sigma}^-_k\hat{\sigma}^+_i=2(\hat{X}_i\hat{X}_k+\hat{Y}_i\hat{Y}_k)$, so that we find our final result
\begin{equation}
    \hat{T}_{ik\uparrow}\to \frac{1}{2}\left (\hat{X}_k\hat{Z}_{k+1}\hat{Z}_{k+2}\cdots \hat{Z}_{i-1}\hat{X}_i+\hat{Y}_k\hat{Z}_{k+1}\hat{Z}_{k+2}\cdots \hat{Z}_{i-1}\hat{Y}_i\right ), 
\end{equation}
with a similar formula for the down spins, given by
\begin{equation}
    \hat{T}_{ik\downarrow}\to \frac{1}{2}\left (\hat{X}_{N+k}\hat{Z}_{N+k+1}\hat{Z}_{N+k+2}\cdots \hat{Z}_{N+i-1}\hat{X}_{N+i}+\hat{Y}_{N+k}\hat{Z}_{N+k+1}\hat{Z}_{N+k+2}\cdots \hat{Z}_{N+i-1}\hat{Y}_{N+i}\right ).
\end{equation}
When the sites are nearest neighbors, there are no $Z$ terms. Using the theorem we just proved, with $\theta=\tfrac{\pi}{2}$, we find that
\begin{equation}
    \braket{\psi|\hat{T}_{ik\uparrow}|\psi}=\braket{\psi|\hat{X}_k\hat{Z}_{k+1}\hat{Z}_{k+2}\cdots\hat{Z}_{i-1}\hat{X}_i|\psi},
\end{equation}
for $i>k$, with a similar formula for $i<k$. Note that the down spin expectation value takes the same form with each index increased by $N$, respectively.

The simplest way to perform this measurement is to rotate the states being measured from the $z$-basis to the $x$-basis (on sites $i$ and $k$ only). Then we simply measure and read off the eigenvalues for each qubit. Finally, they are multiplied together to determine the overall eigenvalue (which is $\pm 1$ for each shot). This is then averaged over many measurements to determine the expectation value.

The rotation used is $\exp\big (-i\tfrac{\pi}{4}\hat{Y}\big )$ for each qubit that is measured in the $x$-direction; this rotation rotates the $z$-axis to the $x$-axis which is the familiar Hadamard gate. 
But, to perform an efficient measurement, we want to measure as many of these expectation values as we can. We describe how to do this next.

If we perform the rotation on every qubit, we can measure the terms with $k=i-1$ and $1\le i\le N$, which is $N-1$ hopping terms in one measurement circuit. What is neat about this simplified (nonentangling) approach is that it allows us to measure overlapping hopping terms because we are in the diagonal basis for the operators. To measure the $k=i-2$ hopping terms, we need two circuits (one starting at $i=2$, the other at at $i=3$)  and can measure approximately half with each circuit (we measure a total of $N-2$ hopping terms with both circuits). Proceeding up to $k=i-j$, with $j< N/2$, we need $j$ measurements, and we measure $N-j-1$ pairs. Once $j\ge N/2$, we need a separate circuit for each measurement. There continues to be $N-j-1$ hopping terms to measure. The total number of terms is then $\sum_{j=0}^{N-2}(N-j-1)=(N-1)^2-\tfrac{1}{2}(N-1)(N-2)=\tfrac{1}{2}N(N-1)$, which is all of the hopping pairs that we have. The number of measurement circuits needed is 
\begin{align}
    M&=\sum_{j=0}^{\text{int}\left[\frac{N-1}{2}\right]}(j+1)+\sum_{j=\text{int}\left[\frac{N+1}{2}\right]}^{N-2}(N-1-j)\nonumber\\
    &=\left (\text{int}\left [\frac{N+1}{2}\right ]\right)^2+(N-1)\left (\frac{N}{2}-\text{int}\left [\frac{N+1}{2}\right ]\right ).
\end{align}
Note that when $N$ is even, this just becomes $(N/2)^2$, and when $N$ is odd, it becomes $((N+1)/2)^2-(N-1)/2$.
As a check, when $N=10$, there are 45 unique fermionic hopping terms for one spin and we need 25 circuits to measure all of them. This agrees with the above formula. In general, we need about half as many measurement circuits as the unique fermionic hopping pairs we have on the lattice. This is reduced, of course, if we do not need to measure all unique hopping pairs in our Hamiltonian. We will discuss this further later in the paper.

There is a second way to measure the hopping, which is the most common way currently used---it does not use the theorem---but instead, it works in a simultaneous eigenbasis of the $\hat{X}\hat{X}$ and $\hat{Y}\hat{Y}$ operators, since they both commute. The eigenbasis that diagonalizes both is the so-called Bell state basis. We have in the case where $k$ is the left qubit and $i$ is the right qubit
\begin{align}
    \ket{\Phi^+}&=\frac{1}{\sqrt{2}}(\ket{00}+\ket{11}):~~\hat{X}\hat{X}\to 1,~~~~~\hat{Y}\hat{Y}\to -1;\\
    \ket{\Phi^-}&=\frac{1}{\sqrt{2}}(\ket{00}-\ket{11}):~~\hat{X}\hat{X}\to -1,~~\hat{Y}\hat{Y}\to 1;\\
    \ket{\Psi^+}&=\frac{1}{\sqrt{2}}(\ket{01}+\ket{10}):~~\hat{X}\hat{X}\to 1,~~~~~\hat{Y}\hat{Y}\to 1;~~\text{and}\\
    \ket{\Psi^-}&=\frac{1}{\sqrt{2}}(\ket{01}-\ket{10}):~~\hat{X}\hat{X}\to -1,~~\hat{Y}\hat{Y}\to -1.
\end{align}
The eigenvalue of $\hat{X}\hat{X}$ is the measured value of $Z_k$ and the eigenvalue of $\hat{Y}\hat{Y}$ is $-Z_kZ_i$, where $Z_k$ is the measured value of 1 or $-1$ on the $k$th qubit, and similarly for $Z_i$, for the given shot. 
The circuit to transform to this basis is simple: (i) first we apply a CNOT using the first qubit as the control and the second qubit as the target, then (ii) we apply a Hadamard to the first qubit. The number of measurement circuits is identical to the first measurement scheme we discussed, except for the set of circuits with $k=i-1$. In this case, we need to run two measurement circuits to provide the entanglement across the qubits connected by the nearest-neighbor hopping. When we have Jordan-Wigner strings present (next-nearest neighbors and further), we also multiply the corresponding $Z$ eigenvalues for each qubit in the string.

One can see that the $\Phi$ states will not contribute to the average, because they change the fermion number in the state after we evaluate $\hat{X}\hat{X}$ or $\hat{Y}\hat{Y}$ onto it. Furthermore the $\Psi$ states have the same eigenvalues for the $\hat{X}\hat{X}$ and $\hat{Y}\hat{Y}$ measurements, which is another way to see why our theorem holds.

In comparing the two approaches, the nonentangling approach uses one fewer measurement circuit and does not require as many entangling gates for the measurement, so it should always perform better, in theory. This becomes more detrimental if the hardware does not have all-to-all connectivity, because some cases may require significant numbers of SWAPs.

This completes the analysis for the quadratic hopping terms. The next section evaluates quartic interaction terms, where the analysis is more complicated.

%%%%%%%%%%%%%%%%%%%%%%%%%%%%%%%%%%%%%%%%%%
\subsection{Application to Quartic Operators}

We now discuss the quartic terms. Since we assume the Hamiltonian is Hermitian, there is no spin-orbit coupling or other terms that can flip spins, and all Hamiltonian matrix elements are real, the most general form involves the following set of terms:
\begin{align}
    &\hat{c}_{i\sigma}^\dagger\hat{c}_{j\sigma}^\dagger\hat{c}_{k\sigma}^{\phantom{\dagger}}\hat{c}_{l\sigma}^{\phantom{\dagger}}+\hat{c}_{l\sigma}^\dagger\hat{c}_{k\sigma}^\dagger\hat{c}_{j\sigma}^{\phantom{\dagger}}\hat{c}_{i\sigma}^{\phantom{\dagger}}~~\text{with}~~i>j,~~k>l,~~\text{and}~~ (ij)\ge (kl),\nonumber\\
    &\hat{c}_{i\sigma}^\dagger\hat{c}_{j\bar{\sigma}}^\dagger\hat{c}_{k\sigma}^{\phantom{\dagger}}\hat{c}_{l\bar{\sigma}}^{\phantom{\dagger}}+\hat{c}_{l\bar{\sigma}}^\dagger\hat{c}_{k\sigma}^\dagger\hat{c}_{j\bar{\sigma}}^{\phantom{\dagger}}\hat{c}_{i\sigma}^{\phantom{\dagger}}~~\text{with}~~ (ik)\ge (jl),
\end{align}
where the lexographic inequality $(ij)\ge (kl)$ means $i>k$, or if $i=k$, then $j\ge l$. This has terms where $\sigma=\uparrow$ and $\sigma=\downarrow$ for the equal spin terms in the top row and  $\sigma=\uparrow$ and $\bar{\sigma}=\downarrow$, and $\sigma=\downarrow$ and $\bar{\sigma}=\uparrow$ for the mixed-spin terms in the bottom row. To construct the Hamiltonian, these terms are multiplied by the real numbers labeled by the same indices and summed over all allowed indices (space and spin).

When we perform the Jordan-Wigner transformation for the same-spin case, we immediately find that
\begin{equation}
    \hat{c}_{i\sigma}^\dagger\hat{c}_{j\sigma}^\dagger\hat{c}_{k\sigma}^{\phantom{\dagger}}\hat{c}_{l\sigma}^{\phantom{\dagger}}+h.c.\to -\frac{1}{16}\hat{\sigma}_i^-\hat{Z}_{i-i}\hat{Z}_{i-2}\cdots\hat{Z}_{j+1}\sigma_j^-\hat{\sigma}_k^+\hat{Z}_{k-1}\hat{Z}_{k-2}\cdots\hat{Z}_{l+1}\hat{\sigma}_l^++h.c.,
\end{equation}
where we neglected the spin degree of freedom of the fermions for clarity. Note that because we have no specific ordering relation between $j$ and $k$ (and even $l$), it is not a simple exercise to simplify the Jordan-Wigner string further in this representation. The minus sign came from the fact that $\hat{Z}_j\hat{\sigma}_j^-=-\hat{\sigma}_j^-$. If $j>k$, then the above form properly describes the fully transformed expression. If $k>j$ and $l<j$, then we would have a Jordan-Wigner string running from $i-1$ to $k+1$ and from $j-1$ to $l$, because $\hat{Z}^2=\mathbb{I}$.

Converting to the $\hat{X}$ and $\hat{Y}$ operators, and ignoring the Jordan-Wigner strings which will commute with these operators (and possible sign changes if the $Z$-string from one pair overlaps with the $Z$-string of the other pair) , we find
\begin{align}
    \braket{\psi|\hat{c}_{i\sigma}^\dagger\hat{c}_{j\sigma}^\dagger\hat{c}_{k\sigma}^{\phantom{\dagger}}\hat{c}_{l\sigma}^{\phantom{\dagger}}+h.c.|\psi}&\to-\frac{1}{16}\braket{\psi|(\hat{X}_i-i\hat{Y}_i)(\hat{X}_j-i\hat{Y}_j)(\hat{X}_k+i\hat{Y}_k)(\hat{X}_l+i\hat{Y}_l)+h.c.|\psi}\nonumber\\
    &=-\frac{1}{8}\bra{\psi}\big(\hat{X}_i\hat{X}_j\hat{X}_k\hat{X}_l+\hat{Y}_i\hat{Y}_j\hat{Y}_k\hat{Y}_l-\hat{X}_i\hat{X}_j\hat{Y}_k\hat{Y}_l-\hat{Y}_i\hat{Y}_j\hat{X}_k\hat{X}_l\nonumber\\
    &~~~~~~~~~~~~\hat{X}_i\hat{Y}_j\hat{X}_k\hat{Y}_l+\hat{Y}_i\hat{X}_j\hat{Y}_k\hat{X}_l+\hat{X}_i\hat{Y}_j\hat{Y}_k\hat{X}_l+\hat{Y}_i\hat{X}_j\hat{X}_k\hat{Y}_l\big)\ket{\psi},
\end{align}
which involves eight independent terms (note that the overall sign may change due to overlapping Jordan-Wigner strings). Using the nonentangled measurement scheme requires four separate measurements for each of these terms after using the theorem with $\theta=\tfrac{\pi}{2}$. This says, schematically, $XXXX=YYYY$, $XXYY=YYXX$, $XYXY=YXYX$, and $XYYX=YXXY$, where the schematic notation implies we are taking an expectation value of the corresponding operators---we drop the bras and kets for simplicity here. This indicates that we would need four separate measurements for each term. But, things simplify further if we use the theorem with $\theta=\tfrac{\pi}{4}$, where $\cos\theta=\sin\theta=\tfrac{1}{\sqrt{2}}$. Then we have that
\begin{align}
    XXXX\to&\frac{1}{4}\Big (XXXX+YYYY+XXYY+YYXX+XYXY+YXYX+XYYX+YXXY\nonumber\\
    &+XYYY+YXYY+YYXY+YYYX+YXXX+XYXX+XXYX+XXXY\Big)\nonumber\\
    &=\frac{1}{2}\Big(XXXX+XXYY+XYXY+XYYX\Big),
\end{align}
after using the theorem with $\theta=\tfrac{\pi}{2}$ on the right hand side to remove many of the terms. Now, this says that $XXXX$ is equal to the expression on the right. If we rearrange, we find that
\begin{equation}
    XXXX-XXYY=XYXY+XYYX,
\end{equation}
which we want to emphasize holds for the expectation values only.
This allows us to reduce the measurements to just two independent measurements. So, we find 
\begin{equation}
    \braket{\psi|\hat{c}_{i\sigma}^\dagger\hat{c}_{j\sigma}^\dagger\hat{c}_{k\sigma}^{\phantom{\dagger}}\hat{c}_{l\sigma}^{\phantom{\dagger}}+h.c.|\psi}\to-\frac{1}{2}\braket{\psi|(\hat{X}_i\hat{X}_j\hat{X}_k\hat{X}_l-\hat{X}_i\hat{X}_j\hat{Y}_k\hat{Y}_l)|\psi},
\end{equation}
where we dropped the Jordan-Wigner strings for simplicity (and there may be a sign change as well). 
For cases where the indices are close enough to each other along the Jordan-Wigner chain, one can perform multiple measurements, which we will discuss in more about in the next section.

\begin{table*}
\caption{Joint eigenstates of $\hat{X}\hat{X}\hat{X}\hat{X}$, $\hat{Y}\hat{Y}\hat{Y}\hat{Y}$, $\hat{X}\hat{X}\hat{Y}\hat{Y}$, $\hat{Y}\hat{Y}\hat{X}\hat{X}$, $\hat{X}\hat{Y}\hat{X}\hat{Y}$, $\hat{Y}\hat{X}\hat{Y}\hat{X}$, $\hat{X}\hat{Y}\hat{Y}\hat{X}$, and $\hat{Y}\hat{X}\hat{X}\hat{Y}$. The states are expanded in terms of the four qubits in the computational ($z$-) basis.} 
    \label{tab:1}
    \renewcommand{\arraystretch}{1.5}
    \begin{tabular}{c | c | c | c}
    \hline
    \textbf{State Label}&\textbf{State}&\textbf{State Label}&\textbf{State}\\
    \hline
    $\ket{\Psi_1^\pm}$&$\tfrac{1}{\sqrt{2}}\big(\ket{0000}\pm\ket{1111}\big)$&$\ket{\Psi_5^\pm}$&$\tfrac{1}{\sqrt{2}}\big(\ket{1000}\pm\ket{0111} \big)$\\
    $\ket{\Psi_2^\pm}$&$\tfrac{1}{\sqrt{2}}\big( \ket{0011}\pm \ket{1100}\big)$&$\ket{\Psi_6^\pm}$&$\tfrac{1}{\sqrt{2}}\big(\ket{0100}\pm\ket{1011} \big)$\\
    $\ket{\Psi_3^\pm}$&$\tfrac{1}{\sqrt{2}}\big( \ket{0101}\pm\ket{1010} \big)$&$\ket{\Psi_7^\pm}$&$\tfrac{1}{\sqrt{2}}\big(\ket{0010}\pm\ket{1101} \big)$\\
    $\ket{\Psi_4^\pm}$&$\tfrac{1}{\sqrt{2}}\big( \ket{0110}\pm\ket{1001} \big)$&$\ket{\Psi_8^\pm}$&$\tfrac{1}{\sqrt{2}}\big(\ket{0001}\pm\ket{1110} \big)$\\
    \hline
   \end{tabular}
\end{table*}

Similar to the quadratic terms, we can also measure all eight elements with one circuit using an entangled basis. It is more complicated than the Bell state basis. We need to find a set of states that diagonalize all of these eight Pauli strings (these eight strings all commute with each other because they all have an even number of like Pauli operators on the different sites).
The states that diagonalize the system are summarized in Table~\ref{tab:1} and the eigenvalues of the eight operators are summarized in Table~\ref{tab:2}. We label the states with an integer index and a parity. Each state is the sum of just two states. This construction generalizes the Bell states used to simultaneously measure the $\hat{X}\hat{X}$ and $\hat{Y}\hat{Y}$ operators for the quadratic terms.

\begin{table*}[h]
\caption{Eigenvalues of the operators $\hat{X}\hat{X}\hat{X}\hat{X}$, $\hat{Y}\hat{Y}\hat{Y}\hat{Y}$, $\hat{X}\hat{X}\hat{Y}\hat{Y}$, $\hat{Y}\hat{Y}\hat{X}\hat{X}$, $\hat{X}\hat{Y}\hat{X}\hat{Y}$, $\hat{Y}\hat{X}\hat{Y}\hat{X}$, $\hat{X}\hat{Y}\hat{Y}\hat{X}$, and $\hat{Y}\hat{X}\hat{X}\hat{Y}$ for the 16 eigenstates found in Table~\protect{\ref{tab:1}}.} 
    \label{tab:2}
    \renewcommand{\arraystretch}{1.25}
    \begin{tabular}{c | c c c c c c c c}
    \hline
    \textbf{State Label}&$\hat{X}\hat{X}\hat{X}\hat{X}$&$\hat{Y}\hat{Y}\hat{Y}\hat{Y}$&$\hat{X}\hat{X}\hat{Y}\hat{Y}$&$\hat{Y}\hat{Y}\hat{X}\hat{X}$&$\hat{X}\hat{Y}\hat{X}\hat{Y}$&$\hat{Y}\hat{X}\hat{Y}\hat{X}$&$\hat{X}\hat{Y}\hat{Y}\hat{X}$&$\hat{Y}\hat{X}\hat{X}\hat{Y}$\\
    \hline
    $\ket{\Psi_1^\pm}$&$\pm1$&$\pm1$&$\mp1$&$\mp1$&$\mp1$&$\mp1$&$\mp1$&$\mp1$\\
    $\ket{\Psi_2^\pm}$&$\pm1$&$\pm1$&$\mp1$&$\mp1$&$\pm1$&$\pm1$&$\pm1$&$\pm1$\\
    $\ket{\Psi_3^\pm}$&$\pm1$&$\pm1$&$\pm1$&$\pm1$&$\mp1$&$\mp1$&$\pm1$&$\pm1$\\
    $\ket{\Psi_4^\pm}$&$\pm1$&$\pm1$&$\pm1$&$\pm1$&$\pm1$&$\pm1$&$\mp1$&$\mp1$\\
    $\ket{\Psi_5^\pm}$&$\pm1$&$\mp1$&$\mp1$&$\pm1$&$\mp1$&$\pm1$&$\mp1$&$\pm1$\\
    $\ket{\Psi_6^\pm}$&$\pm1$&$\mp1$&$\mp1$&$\pm1$&$\pm1$&$\mp1$&$\pm1$&$\mp1$\\
    $\ket{\Psi_7^\pm}$&$\pm1$&$\mp1$&$\pm1$&$\mp1$&$\mp1$&$\pm1$&$\pm1$&$\mp1$\\
    $\ket{\Psi_8^\pm}$&$\pm1$&$\mp1$&$\pm1$&$\mp1$&$\pm1$&$\mp1$&$\mp1$&$\pm1$\\
    \hline
   \end{tabular}
\end{table*}

The circuit to put the system into this basis is the inverse of creating a so-called NOON state from the $\ket{0000}$ state. It is a set of three CNOTs that use qubit 0 as control and qubits 3, 2, and 1, in turn as the target, followed by a Hadamard on qubit 0. This is a fairly simple entangling circuit to employ.

The measurement on the four qubits will produce a state $\ket{ijkl}$ that is measured. That state is the initial state, from which the states in Table~\ref{tab:1} are prepared after applying a Hadamard on qubit zero followed by three CNOTS, all using the zeroth qubit as a control and the first, second, and third qubits as targets. Once the state is decoded, then Table~\ref{tab:2} lists the eigenvalue for the corresponding operator. Because the circuit will require a potentially large number of swaps to carry out the entangling gates (when on hardware that does not have all-to-all connectivity), the actual cost for the entangled measurement can be significant, in terms of loss of fidelity. 

Because the nonentangled circuit always requires two measurements, while the entangled circuit requires only one, it seems like the entangled circuit is always better. This is certainly true in the general case, unless the entangling operation causes too much error in the result. But, in cases where the operator has no Jordan-Wigner strings, then the nonentangled circuit, will allow a number of measurements to be carried out at once, similar to what happened for the quadratic operators. We will not go into the details for how this works here, nor will we go into the counting of the number of measurement circuits, as it becomes quite involved.

%%%%%%%%%%%%%%%%%%%%%%%%%%%%%%%%%%%%%%%%%%
\section{Discussion}
\subsection{State Preparation in Physics}

The most common system to study in Physics is the Hubbard model~\cite{hubbard}, which is of the form
\begin{equation}
    \hat{H}=-\sum_{ij=1}^N\sum_\sigma t_{ij}\hat{c}_{i\sigma}^\dagger\hat{c}_{j\sigma}^{\phantom{\dagger}}+U\sum_{i=1}^N\hat{c}_{i\uparrow}^\dagger\hat{c}_{i\uparrow}^{\phantom{\dagger}}\hat{c}_{i\downarrow}^\dagger\hat{c}_{i\downarrow}^{\phantom{\dagger}},
\end{equation}
where $t_{ij}$ is a real, symmetric matrix. A common problem people wish to solve is a ground state preparation problem. A number of different approaches have been proposed for this that follow a variational quantum eigensolver approach, but we find a recent approach that employs just hopping terms along the Jordan-Wigner chain and the double occupancy terms to be on of the most efficient approaches~\cite{he_efficient_2025}.

In working on implementing the algorithm, one needs to measure the energy and derivatives of the energy. The latter is often handled by using the parameter shift rule, so that one essentially measures the energy, or energy-like objects for everything.

Here, the nontrivial measurement circuits are for the quadratic terms, which are best measured using the nonentangled circuits. One will need a number of measurement circuits for lattices in two and three dimensions, because the hopping matrix will connect sites that are not adjacent along the Jordan-Wigner chain. But, for cases where the lattice is periodic and satisfies sufficient rotational and reflection symmetry in the point group, then we have that all hopping matrix elements have the same expectation value, so just the measurement along the Jordan-Wigner chain will suffice. The double occupancy term can just be measured directly in the computational basis and requires only one circuit. 

In the most common case of a periodic lattice with nontrivial point-group symmetry, the method we outlined here is the most efficient approach. Not just because the hopping and the double occupancy can each be measured with just one measurement circuit, but also because we can combine the results from $N-1$ hopping terms into a single measurement, which provides additional statistics at no extra cost. 

Note further, that one needs to properly address the shot budget for a given calculation. While one might naively want to measure both hopping and double occupancy with the same number of shots, one might change to a different number if one term or the other is needed with higher accuracy. This allows one to then reduce the number of shots for one of the expectation values and increase for the other. We will not go into an analysis for how this is done, but just state that one of the ideas for how to make these choices is by minimizing the variance for each measurement, and then for their sum.

This approach also works for a case with reduced symmetry (for example, if one has disorder), just with more measurement circuits needed for the hopping terms. We will not go into further details about how that works here.

\subsection{State Preparation in Chemistry}

While ground state preparation in Chemistry resembles the same problem in Physics, the details are quite different. In Physics, the Hamiltonian, is usually fairly sparse, such as the Hubbard model, while in Chemistry, there are many, many more interaction terms. But, as one might expect, the different terms have different sizes, and one needs to take this into account when trying to determine a measurement protocol for these systems.

Because the interaction terms are quartic operators, if we have $2N$ spin orbitals, we have on the order of $N^4$ interaction terms. Even for a small system, most of the measurements will be for these terms. The sizes of these terms depends significantly on how they are formed. The hopping terms and the Coulomb repulsion terms within a spatial orbital are the most important, with energies on the order of a Hartree, or a large fraction of it. The next important are the two-site integrals, which involve density-density Coulomb interactions and exchange. Then come the three and four-site integrals, which typically are substantially smaller.

This hierarchy of terms brings up a clear measurement strategy. We use nonentangled circuits for the hopping terms and the direct Coulomb terms (within a spatial orbital) and we use entangled circuits for everything else. We then balance the total shot budget according to the size of the term and the accuracy needed. This will reserve most shots to the hopping and largest interaction terms, whereas the smaller interaction terms will be measured much less frequently. Performing a detailed analysis to determine how many times to measure each term is an analysis that requires one to specify the details of the molecule as well as the layout and connectivity of the hardware and the mapping of spin orbitals to qubits. There are many places where one needs to think carefully to provide the mapping in the most efficient fashion. We do not discuss those details further here, because they require concrete details about the specific molecules and how we would work with them.

In any case, we expect that accuracy will improve by employing this hidden symmetry, which allows for more efficient and more accurate implementation of measurement circuits.

%\begin{figure}[H]
%\isPreprints{\centering}{} % Only used for preprints
%\includegraphics[width=4.0 cm]{logo-mdpi}
%\caption{This is a figure. Schemes follow the same formatting.\label{fig1}}
%\end{figure}   

%\begin{table}[H] 
%\small % Change table font size
%\caption{This is a table caption. Tables should be placed in the main text near to the first time they are~cited.\label{tab1}}
%\isPreprints{\centering}{} % Only used for preprints
%\begin{tabularx}{\textwidth}{CCC}
%\toprule
%\textbf{Title 1}	& \textbf{Title 2}	& \textbf{Title 3}\\
%\midrule
%Entry 1		& Data			& Data\\
%Entry 2		& Data			& Data \textsuperscript{1}\\
%\bottomrule
%\end{tabularx}

%\noindent{\footnotesize{\textsuperscript{1} Tables may have a footer.}}
%\end{table}

%%%%%%%%%%%%%%%%%%%%%%%%%%%%%%%%%%%%%%%%%%
\section{Conclusions}

In this work, we have explored a hidden U(1) symmetry that appears after one has performed a Jordan-Wigner transformation from fermions to spins. Because the Pauli $X$ and $Y$ matrices are the ``real'' and ``imaginary'' parts of the creation and annihilation operators, this symmetry should also be present in treatments that employ Majorana fermions. But, we do not go into further details here about that.

For quadratic operators, the nonentangled measurement circuits that exploit this symmetry are significantly more efficient than the entangled counterparts, and because the circuits use less entangling operations, they should be more accurate to implement. For quartic operators, the nonentangled approach is usually less efficient, except for the most important quartic operators, which involve density-density interactions within the same spatial orbital. But the precise analysis for how to achieve a savings using these circuits depends significantly on the qubit layout in the hardware.

We believe this symmetry will have additional applications within quantum information in the problem of simulating fermionic systems. We look forward to future work that will exploit this symmetry further.

%%%%%%%%%%%%%%%%%%%%%%%%%%%%%%%%%%%%%%%%%%
\vspace{6pt}

%%%%%%%%%%%%%%%%%%%%%%%%%%%%%%%%%%%%%%%%%%
\noindent\textbf{Author Contributions:}{ Conceptualization, J.K.F.; methodology, J.K.F; validation, G.D.; formal analysis, J.K.F. and G.D.; investigation, J.K.F. and G.D.;  writing---original draft preparation, J.K.F.; writing---review and editing,J.K.F. and G.D.; supervision, J.K.F.;  funding acquisition, J.K.F. All authors have read and agreed to the published version of the manuscript.}\\

\noindent\textbf{Funding:}{ G.D. was funded by the Department of Energy, Office of Advanced Scientific Computing Research, Accelerated Research in Quantum Computing program  under Grant No. DE-SC0025483 for the work described in the author contributions. J.K.F. was funded by the National Science Foundation grant number CHEM-2154671  and by the McDevitt bequest at Georgetown University for the work described in the author contributions. The APC was funded by MDPI. }\\

\noindent\textbf{Data availability:}{ No data was created for this work.}

\noindent\textbf{Conflicts of Interest:}{ The authors declare no conflicts of interest.}

\bibliography{jw_bib.bib}

\end{document}